# ASTEROID PHOTOMETRY

**Jian-Yang Li**
*Planetary Science Institute*

**Paul Helfenstein**
*Cornell University*

**Bonnie J. Buratti**
*California Institute of Technology, Jet Propulsion Laboratory*

**Driss Takir**
*Ithaca College*
*Planetary Science Institute*

**Beth Ellen Clark**
*Ithaca College*

Abstract

Asteroid photometry has three major applications: providing clues about asteroid surface physical properties and compositions, facilitating photometric corrections, and helping design and plan ground-based and spacecraft observations. The most significant advances in asteroid photometry in the past decade were driven by spacecraft observations that collected spatially resolved imaging and spectroscopy data. In the mean time, laboratory measurements and theoretical developments are revealing controversies regarding the physical interpretations of models and model parameter values. We will review the new developments in asteroid photometry that have occurred over the past decade in the three complementary areas of observations, laboratory work, and theory. Finally we will summarize and discuss the implications of recent findings.

# 1. INTRODUCTION

## 1.1 Importance of planetary photometry

Planetary photometry concerns how the brightness of a planetary surface depends on the illumination and observing geometry. Fig. 1 shows the scattering geometry and definition of terms that will be important for the rest of this chapter. The angle between the light source and local surface normal is incidence angle, $I$; the angle between the observer and the local surface normal is emission angle, $e$; and the angle between light source and observer, as seen from the object, is phase angle, $\alpha$. The dependence of the brightness of the surface on $(i, e, \alpha)$ is determined by the optical and mechanical properties of the surface.

[Fig. 1]

Why is asteroid photometry important? We broadly divide the applications of planetary photometry, especially with spatially resolved data, into three major aspects.

First, it provides clues about the physical properties and compositions of asteroidal surfaces. How the reflectance of a particulate surface responds to the illumination and viewing geometry is fully determined by the optical and mechanical properties of a surface, such as particle size, porosity, roughness, complex refractive index, etc. The ultimate goal of photometric modeling is to derive these fundamental properties of a surface by measuring the reflectance under various light scattering geometries. Despite the development of various theoretical models, empirical or semi-physical, over the past several decades, we are still far from being able to determine all surface physical conditions from photometric observations and modeling of asteroids. The major reason is that real asteroidal surfaces have complex physical conditions that are difficult to fully parameterize. We still do not fully understand the light scattering effects of some physical conditions, such as roughness, porosity, etc. Therefore it is nearly impossible to find exact, unique solutions of radiative transfer equations that describe real asteroidal surfaces. Some assumptions and approximations have to be made in order to derive practical analytical models, such as the Hapke model (Hapke, 1981; 1984; 1986; 2002; 2008; 2012a), the Shkuratov model (Shkuratov et al., 1999), etc. Observationally, due to the limited observing geometries available for ground-based and spacecraft observations to any asteroid, it is often difficult to fully constrain a photometric model for a particular object.

Second, because the spectral reflectance of a surface depends on illumination and viewing geometry, it is necessary to correct observations (imaging or spectral) to a common (reference) geometry in order to make comparisons between different areas on an asteroid, between different asteroid surfaces, and with laboratory measurements. Photometric modeling provides such a means for this purpose. For example, phase reddening is a common phenomenon that was first reported by Gehrels et al. (1964) and has been recognized on many asteroids and for a long time (e.g., Clark et al., 2002 and references therein; Reddy et al., 2012a; Li et al., 2013a; also see Section 4.1.2) that causes the spectrum of a surface to have redder slopes and stronger spectral absorptions at higher phase angles. One has to correct for this in order to make meaningful mineralogical interpretations of reflectance spectra. Other examples include mosaicking albedo and color maps of the whole surface of an asteroid from spacecraft data taken at vastly different illumination geometries (e.g., Li et al., 2013a). In addition, photometric correction makes it possible to compare asteroid observations with laboratory measurements, which are usually made at a standard geometry. The general procedure of photometric correction is to fit a photometric model, $r_{model}(i, e, \alpha)$, to the observations, $r(i, e, \alpha)$, then follow the Eq.1 below to calculate the corrected reflectance under a reference geometry $r(i_0, e_0, \alpha_0)$:

$$r(i_0, e_0, \alpha_0) = \frac{r(i_{measure}, e_{measure}, \alpha_{measure})}{r_{model}(i_{measure}, e_{measure}, \alpha_{measure})} \times r_{model}(i_0, e_0, \alpha_0) \qquad (1)$$

where ($i_{measure}$, $e_{measure}$, $\alpha_{measure}$) is the scattering geometry of measured reflectance. The most commonly used reference geometries for correction are ($i_0$, $e_0$, $\alpha_0$) = (0º, 0º, 0º), which corresponds to the normal reflectance, and ($i_0$, $e_0$, $\alpha_0$) = (30º, 0º, 30º), which is a common laboratory setting.

The third application of photometry (that we consider) is to predict the reflectance of an asteroidal surface at arbitrary illumination and viewing geometries for designing and planning observations, especially for asteroid spacecraft exploration missions. Spacecraft generally observe target asteroids at substantially different viewing geometries from ground-based observers. It is therefore necessary to extrapolate or interpolate the photometric measurements from the ground to calculate the nominal brightness of a target for planning purposes. For spatially resolved images, it is necessary to know the dependence of reflectance on local scattering geometry. This practical application is important for essentially all spacecraft missions with imaging or spectroscopy components.

**1.2 Scope of this chapter**

Except for several of the largest asteroids, most asteroids appear to be point sources when observed from the ground, and the dependence of reflectance on local scattering geometry is unavailable. Therefore, up until the first asteroid flyby of (951) Gaspra by the Galileo spacecraft *en route* to the Jupiter system in 1991 (Belton et al., 1992), asteroid photometry focused only on studying the dependence of total brightness of an object on solar phase angle, the so-called phase function (Fig. 2). In order to retrieve physical properties of the surface from photometric phase functions, empirical and experimental models have been developed (cf. Muinonen et al., 2002 and references therein). For example, it has long been recognized that the brightness of dark objects generally decreases with phase angle faster than bright objects, resulting in relatively steeper slopes of their phase functions, a phenomenon attributed to relatively stronger multiple scattering in brighter objects (cf. Bowell and Lumme, 1979 and references therein). In addition, higher surface roughness also contributes to the steeper slope of a phase function (e.g., Veverka, 1971), although it is not possible to entirely disentangle the effects of multiple scattering and surface roughness from a disk-integrated phase function.

[Fig. 2]

The most significant development in observational studies of the photometric properties of asteroids over the past decade has come from the availability of high-resolution, spatially resolved images of the surfaces of many more asteroids from flyby and rendezvous missions with asteroids (e.g., Li et al., 2006; Kitazato et al., 2008; Spjuth et al., 2012; Magrin et al., 2012; Li et al., 2013a; Schröder et al., 2013). From these data, we can observe how the reflectance of an asteroidal surface depends on local scattering geometries. This additional information is more powerful for retrieving the physical properties of a surface than the disk-integrated solar phase function only. The most notable advantage is to resolve the ambiguity between fitting the surface roughness and the phase function, because the effect of roughness is especially important at phase angles >40° (Helfenstein, 1988).

In this chapter, we will focus on disk-resolved photometric studies of asteroids based on photometric theories and high-resolution data returned by asteroid exploration missions. Photometric modeling based primarily on ground-based photometric surveys have been thoroughly reviewed in previous *Asteroids* books of this series (Bowell and Lumme, 1979; Bowell et al., 1989; Helfenstein and Veverka, 1989; Muinonen et al., 2002). It is important to keep in mind that physical laws dictate that the photometric signatures from planetary surfaces necessarily correlate with polarimetric signatures. In this chapter we strictly limit the scope of our discussion to asteroid photometry, and readers are referred to a companion chapter by Belskaya et al. for a comprehensive review on asteroid polarimetry. In Section 2, we will review the basic concepts of photometric measurements and models, the most recent developments in theoretical models, and the controversy about several analytical models. In Section 3, we will summarize observational results reported over the past decade, focusing on disk-resolved photometry based on spacecraft observations. We will then discuss applications and implications of photometric modeling to the study of general asteroid properties in Section 4. In Section 5, we provide a summary and a perspective on the future of asteroid photometry for the next decade to come.

## 2. OVERVIEW OF THEORIES

### 2.1 Basic Concepts

The fundamental quantity of light scattering characteristics of a surface is reflectance. Several different quantities of reflectance and albedo exist under various illumination and observing conditions. Here we follow the definitions and conventions described in Hapke (2012b). Generally, reflectance is defined by the ratio of scattered radiance (or intensity) to incident irradiance (or flux), while albedo is defined with respect to an ideal surface that scatters all incident light isotropically (or a perfectly scattering Lambert surface). Reflectance quantities use two adjectives as prefixes to specify the collimation of incident light and the measurement conditions for scattered light, such as *directional-directional* reflectance, *directional-hemispherical* reflectance, etc. In the case where the two adjectives are the same, a prefix *bi-* is used, e.g., *bi*directional reflectance. The most commonly used quantities in light scattering theories and measurements are listed in Table 1. Here we describe a few of the most important quantities.

[Table 1]

*Bidirectional reflectance*, $r$, is an idealized quantity, because the incident irradiance, $J$, is assumed to be strictly collimated (Fig. 1). For observations of most asteroids, the apparent angular size of the Sun is <0.5° (at >1 AU from the Sun), so the collimation assumption for incident irradiance is a good approximation except for at near zero phase angle, where the finite size of the Sun rounds off the peak. The measurements of scattered intensity are made at the pixel scale of the camera or spectrometer, which is typically less than a few tens of milliradians per pixel. $J$ has a unit of [W m$^{-2}$] or [W m$^{-2}$ per unit wavelength or per unit frequency]). The scattered radiance, $I$, has a unit of [W m$^{-2}$ sr$^{-1}$] or [W m$^{-2}$ sr$^{-1}$ per unit wavelength or unit frequency]. Therefore, bidirectional reflectance has a unit of [sr$^{-1}$].

*Radiance factor* (or RADF), $R$, and *reflectance factor* (or REFF) are often used in laboratory measurement by ratioing the reflected light from the sample to that from a reference surface, which is usually close to a Lambert disk. A Lambert surface has a bidirectional reflectance $r_L(0, e, e) = 1/\pi$ in any direction, with a unit of [sr$^{-1}$]. Therefore, both RADF and REFF are dimensionless. Note that non-isotropic scattering, especially a phenomenon known as the "opposition effect" near zero phase (a non-linear increase of reflectance as phase angle approaches 0º; see Section 2.3), can make a surface brighter than a perfectly scattering Lambert surface, producing RADFs greater than unity.

RADF is equivalent to the commonly used but often confusing notation $I/F$, which is usually annotated in the literature as "$I$ is the scattered radiance, and $\pi F$ is the incident solar irradiance". Note, however, that the $\pi$ here originates from the division of the bidirectional reflectance quantity of a perfectly scattering Lambert surface, which is $1/\pi$ with a unit of [sr$^{-1}$]. *As such, the $\pi$ actually has a unit of [sr].* Therefore, $F$ has a unit of radiance rather than irradiance, making $I/F$ dimensionless.

Other important concepts include geometric albedo (or physical albedo) and Bond albedo (or spherical albedo). Similar to RADF, *Geometric albedo*, $A_P$, is defined with respect to a perfectly scattering Lambert surface. The use of geometric albedo simplifies the modeling of the disk-integrated brightness of an object at any phase angle, which can now be expressed as the product of $A_P$ and its disk-integrated phase function, $\Phi(\alpha)$, normalized to unity at zero phase angle. Note that, similar to RADF, for extremely bright and strongly backscattering objects, the geometric albedo can approach or exceed unity. E.g., the geometric albedo of Enceladus is 1.38, Tethys 1.23, and Dione 1.00 (Verbiscer et al., 2007).

*Bond albedo*, $A_B$, (also known as the spherical Bond albedo) is a key quantity to measure the ability of an object to absorb incident energy, therefore critical for understanding energy balance and volatile transport on a planetary body. By definition, Bond albedo cannot exceed unity. Since Bond albedo is an integrated quantity of the disk-averaged reflectance, it can be expressed as,

$$A_B = A_P q \qquad (2)$$

where $q$ is the phase integral, defined as,

$$q = 2 \int_0^\pi \Phi(\alpha) \sin \alpha \, d\alpha \qquad (3)$$

Important for thermal modeling is the bolometric Bond albedo, which is the average Bond albedo over wavelength, weighted by the solar spectrum, $F_\odot(\lambda)$,

$$A_B = \frac{\int_0^\infty A_B(\lambda) F_\odot(\lambda) \, d\lambda}{\int_0^\infty F_\odot(\lambda) \, d\lambda} \qquad (4)$$

Because the solar spectrum peaks at about 500 nm with about half of the total flux in the visible wavelengths, the *V*-band Bond albedo is often taken as an approximation to the bolometric Bond albedo for asteroids.

## 2.2 Empirical models

Sophisticated modern photometric models need to describe two types of photometric data: "whole-disk" or "disk-integrated" observations and "disk-resolved" or "surface-resolved" reflectance measurements when they are available; the latter most often obtained from spacecraft borne instruments. Disk-resolved photometric measurements provided a new ability to detect the photometric effects of physical phenomena like macroscopic surface texture much more reliably and unambiguously than could be achieved with whole-disk data.

Surface-resolved photometric models that are applied to asteroid observations seek to relate the local scene viewing and illumination geometry to the radiance factor, RADF, or, more simply, $R$. In the simplified treatment of the empirical equations to model $R$, such as that of Lambert (1760), Minnaert (1941), the dependence of reflectance on $i$ and $e$, usually called the *disk-function* or *limb-darkening function*, $d(i, e)$, is often separated from the dependence on phase angle, called the *surface phase function*, $f(\alpha)$. The RADF of a surface is expressed as,
$$R = d(i, e) f(\alpha) \qquad (5)$$
Sometimes a scaling factor is added to Eq. 5, so that the surface phase function can be normalized, e.g., to unity at zero phase angle. Generally, the disk-function is affected by the amount of multiple scattering (therefore albedo) and surface roughness. The phase function includes the effects of single scattering phase function, opposition surge (see Section 2.3), roughness, and multiple scattering. Historically, this separation is a result of the lack of surface-resolved data before spacecraft missions, where the only available geometric variable was phase angle. Modern photometric theories indicate that when multiple scattering is not significant (i.e., for relatively dark surfaces), the disk-function and phase function can be separated in functional forms.

The most commonly used empirical photometric models are listed in Table 2. Most empirical photometric models specify the disk-function, including the Lambert model (Lambert, 1759), the Lommel-Seeliger (LS) model (Seeliger, 1987), the Minnaert model (Minnaert, 1941), and the lunar-Lambert model (e.g., Buratti and Veverka, 1983; McEwen, 1991; 1996), while leaving the phase function implicit or unspecified. These models can describe surfaces with a wide range of different albedos. Generally, high reflectance objects with geometric albedo close to or greater than unity are well described by the Lambert model, although essentially no asteroids scatter light following Lambert law because the majority have albedos <0.5 (Masiero et al., 2011). On the other hand, low reflectance objects with geometric albedo <0.2 generally follow an LS scattering law. All primitive asteroids (C-, D-, P-types) and most S-type asteroids are in this category. For the highest albedo S-type asteroids and high albedo asteroid classes such as V- and E-types, the hybrid lunar-Lambert model is a good disk-function. The higher the albedo, the greater the partition to the Lambertian term. The Minnaert model can also describe surfaces with a wide range of albedos. The two Minnaert parameters $A_M$ and $k$ depend on phase angle. For a dark surface where multiple scattering is negligible, $k$=0.5; for a Lambert scattering law, $k$=1; and in general the higher the albedo of the surface, the higher the value of $k$ (McEwen, 1991). Buratti (1984) showed that *k*=0.5 applied to normal reflectances ≤ 0.6, which include most asteroids.

[Table 2]

The surface phase function is often expressed in various empirical forms in the literature (Table 3), including the combination of an exponential term and a polynomial term (Hillier et al., 1999; Buratti et al., 2011), or a linear-exponential function (Piironen, 1994; Kaasalainen et al., 2001, 2003; Muinonen et al., 2009), or a linear function in magnitude scale (-2.5log(*r*)) (Li et al., 2009; 2013b), or exponentials and a polynomial in a magnitude scale (Takir et al., 2015).

[Table 3]

Disk-integrated photometric models, as the name implies, compute the whole-disk brightness of an asteroid at a given apparition by integrating the predicted surface-resolved radiance factors over the fraction of the surface that is both illuminated and visible to the observer. A *disk-integrated phase function*, $\Phi(\alpha)$, can be calculated as,

$$\Phi(\alpha) = \frac{f(\alpha)}{\pi} \int_{\Omega(i,v)} d(i,e) \mu \, d\Omega. \qquad (6)$$

$\Phi(\alpha)$ is generally normalized to unity at zero phase angle (note that Eq. 6 is not normalized), and constrained by observational data spanning a range of phase angles. The disk-integrated phase function is therefore different from the surface phase function by including the integral term in Eq. 6 that also depends on phase angle. For simplicity and in the absence of information about the actual shape of an asteroid, the whole-disk brightness is often analytically evaluated for a sphere that has the same equivalent dimensions assumed for the asteroid. The disk-integrated phase functions calculated based on the commonly used photometric models are listed in Table 4.

[Table 4]

Another form of empirical disk-integrated photometric model simply parameterizes the total brightness of an object measured at a range of phase angles, such as the IAU H-G model (next section). In addition, the three-parameter model of Shevchenko (1996) has recently been used to describe the opposition effect of 4 Vesta (Hasegawa et al., 2014). This model, like the IAU H-G one, is expressed in terms of reduced magnitude:

$$M(\alpha) = M(0°) + \frac{a\alpha}{1+\alpha} + b\alpha \qquad (7)$$

where the model parameters are $M(0°)$, the reduced magnitude at zero phase; *a* characterizes the opposition effect, and $b_\lambda$ describes the slope of the linear part of the phase curve, analogous to the phase coefficient.

## 2.3 Physically Motivated Models

Physically motivated models (e.g., Hapke, 1963; 1981; 1984; 1986; 2002; 2008, 2012a; Goguen, 1981; Lumme and Bowell, 1981a; 1985; Shkuratov et al., 1999) approximate the solution for radiative transfer from a rough, particulate surface. A typical modern analytical photometric model can be best considered as an assemblage of mathematical components, each of which describes a functional dependence of directional light scattering on a different parameterized quasi-physical property, or a group of related properties that act in concert. The main functional components of typical models rely on (1) a component to describe albedo and directional singly scattered light by an average regolith grain, (2) a component to describe how wave fronts are multiply scattered among and from an aggregate volume of average grains, (3) a parameterized component to describe the effects of macroscopic surface texture on reflected light, and (4) component functions to describe the opposition effect – a conspicuous non-linear surge in brightness with decreasing phase angle that is observed at small phase angles on particulate covered bodies. So complex are the interactions between these components that even the best available analytical models are still in development. They have continued to evolve over a span of at least three decades as laboratory testing and refinement of theory identify new deficiencies and new corrections, respectively.

Among the analytical models used for asteroid work, the Hapke (1981; 1984; 1986; 2002; 2008; 2012a) model, the Shkuratov (1999) model, and the Lumme-Bowell model (1981a; b; Bowell et al., 1989) have seen the widest application to both photometric and spectroscopic data. The Lumme-Bowell (1981a; b) model is described in detail and discussed in *Asteroids II* volume of this series (Bowell et al., 1989) and later discussion was given in *Asteroids III* (Muinonen et al., 2002). In all cases, there are both a surface-resolved expression and a disk-integrated expression in the model. In the case of the Hapke model, there are many variants. So in practice, whole-disk modeling is done by numerical integration over an assumed spherical or triaxial-shape, or a 3-D shape model, any of which is oriented at the appropriate geometry to match the observation. We discuss below recent advances in these models.

*2.3.1 Hapke Models*

The Hapke (1981,1984,1986) model was discussed in detail in the *Asteroids II* volume (Bowell et al., 1989; Helfenstein and Veverka, 1989). Since then, the model has continued to evolve through experimental variants (cf. Helfenstein et al., 1997), diverse choices in the particle phase functions used in the model, and through numerous improvements introduced by Hapke (2002, 2008, 2012a) and others. Hapke (2002) improved his treatment of multiple scattering by grains with anisotropic phase functions and added the coherent backscattering as a mechanism to account for a narrow (less than a few degrees) opposition spike that acts in addition to the shadow hiding mechanism. By far, the most significant improvement is in Hapke (2008), which identifies and corrects a serious error in how the effects of porosity and compaction are treated in nearly all similar models.

[Table 5]

Although Hapke provides analytical expressions for the whole-disk behavior (cf. Hapke, 2012b), because of the many variants of the model in use, whole-disk phase curves predicted by the Hapke model are often obtained by numerical integration of RADF over the illuminated portion of the planetary disk or asteroid shape model.

The current form of Hapke's radiance factor equation can be expressed as

$$R(\mu_0, \mu, \alpha) = K \frac{\varpi_0}{4} \frac{\mu_0'}{\mu_0' + \mu'} \left[ (1 + B_{SH}(\alpha)) P(\alpha) + M\left(\frac{\mu_0'}{K}, \frac{\mu'}{K}, \alpha\right) \right] (1 + B_{CB}(\alpha)) S(\mu_0', \mu', \alpha)$$

(8)

where $\mu_0'$ and $\mu'$ are the cosines, respectively, of the *effective* angles of incidence and emission angles after adjustment for the average slope angle of macroscopic scale topographic roughness (Hapke, 1984), $\alpha$ is the phase angle, and $\varpi_o$ is the average particle single scattering albedo. The function, $S(\mu_o', \mu', \alpha)$ describes the shadowing and occlusion effects of macroscopic roughness, $P(\alpha)$ is the average particle single scattering phase function (also known as the single particle phase function, SPPF), $(1+B_{SH}(\alpha))$ describes the shadow-hiding opposition effect (Hapke, 1986), $M(\mu_o', \mu', \alpha)$ describes multiply scattered light between particles (Hapke, 2002), and $1+B_{CB}(\alpha)$ models the coherent backscatter opposition effect (Hapke, 2002). The porosity coefficient, $K$ was introduced in Hapke (2008) as an explicit adjustable parameter to more realistically model how the areal fraction of light that is blocked by grains within a quasi-continuous particulate medium increases as the particle arrangement converges to a closely-packed state.

*2.3.1.1 Single scattering by average regolith grains*

Light that is scattered once by an average regolith grain is generally described by the product $\varpi_o P(\alpha)$ where $\varpi_o$ is the average particle single scattering albedo and $P(\alpha)$ is known as the average particle single scattering phase function. The latter describes the directional scattering behavior of average regolith grains. Since the first models of Hapke (1963; 1966), a wide variety of particle phase function expressions have been introduced to the Hapke equations (Table 5), many of which are still in use.

Currently, the most often used representation is that from McGuire and Hapke (1995). Hapke (2012a) recently introduced a simplification of the 2PHG, who found from laboratory studies that the *b* and *c* parameters could be defined in terms of one another using an empirical "hockey-stick relation" (Hapke, 2012a) so named because of the shape defined in *b* vs. *c* plots of model fits to laboratory data. The hockey-stick diagram (i.e. plot of the *c* vs. *b*) is particularly useful for interpreting the physical structure of regolith particles from their particle phase functions with respect to the complexity of the particle shape and the density of internal scatterers (cf. McGuire and Hapke, 1995; Souchon et al., 2011).

*2.3.1.2 Opposition Surge: Shadow Hiding Opposition Effect (SHOE)*

The treatment of Hapke (1986) is most widely in current use. This model uses two adjustable parameters, here called $h_{SH}$ and $B_{0,SH}$, to describe the angular width and amplitude of the SHOE, respectively. It takes into account how non-uniform states of compaction with depth can influence the angular width of SHOE and how the transparency of regolith grains affects the strength of SHOE. The angular half-width of the SHOE, $\Delta\alpha_{SH} \approx 2\, h_{SH}$ (in radians). While the Hapke (1986) model can describe SHOE for many different gradients in compaction state and particle size-distributions, most workers assume a uniform distribution of particles and compaction. In this case $h_{SH}$ can be easily related to the regolith porosity, $p$, in terms of the packing factor $\phi = (1-p)$. Then, $h_{SH} = -0.375 \ln(1-\phi)$.

However, in a related paper, Hapke (2008) identified a fundamental error in the way that his own and almost all other radiative transfer approximations for regolith surfaces. Nearly all of these models fail to account for the increase in reflectance seen in laboratory experiments when particulate surfaces are compacted from a more fluffy state. Hapke (2008) corrects for this effect of porosity at the expense of adding an explicit model parameter called the porosity coefficient, $K$, where

$$K = -\frac{\ln(1-1.209\varphi^{2/3})}{1.209\varphi^{2/3}} \quad (9)$$

A less rigorous version of $K$ figured in earlier versions of Hapke's model where it is called the porosity factor. The porosity factor figures in Hapke's (1986) derivation of the SHOE function, which strongly depends on porosity. Helfenstein and Shepard (2011) introduced a correction to the Hapke (1986) SHOE model to make it consistent with Hapke (2008). However, it is important to note that the Hapke (2008) correction is strictly valid only for porosities in the range $0.248<p<1.0$.

*2.3.1.3 Opposition Surge: Coherent-Backscatter Opposition Effect (CBOE)*

Coherent-backscatter is now known to cause the prominent, very narrow phase-curve spike that is observed most often within a few degrees of opposition (Muinonen et al., 2012). How the mechanism operates in the complex architecture of planet, satellite, and asteroid regoliths is still not fully understood[1] (cf. Hapke et al., 2012). In his most recent treatment (Hapke, 2012a), he updates his approximation to include the dependence on the porosity factor,

$$B_{CB}(\alpha) = \frac{B_{0,CB}}{1+1.42K}\left[1+\frac{1-e^{-(1.42K/h_{CB})\tan(\alpha/2)}}{(1/h_{CB})\tan(\alpha/2)}\right] \quad (10)$$

where $B_{0,CB}$ is the amplitude of the CBOE, $h_{CB}=\lambda/4\pi L_T$, $\lambda$ is wavelength, and $L_T$ is the photon transport mean free path length in the medium. The angular half-width of the CBOE in radians is $\Delta\alpha_{CB}=0.36\lambda/2\pi L_T$. While it is explicitly proportional to the spectral wavelength, it is important to note that the transport mean free path length in the medium is often strongly wavelength-dependent. Hence, the angular width of the CBOE in a complex regolith can be difficult to predict.

*2.3.1.4 Multiple-Scattering of Light*

Hapke's (1981, 1984, 1986) model approximated multiple scattering with an analytical simplification of Chandrasekhar's (1960) H-functions for isotropically scattering grains (known as *IMSA*, or the Isotropic Multiple Scattering Approximation). This approximation has long been assumed to be adequate for relatively low to moderate albedo bodies where the contribution of multiply scattered light was small compared with the emitted singly scattered signal. For relatively high-albedo objects where multiple-scattering contributes significantly, some workers replace the approximation with the exact numerical calculation for isotropic scatterers (cf. Verbiscer, 1991) and first-order anisotropic scatterers (cf. Verbiscer and Veverka, 1992; 1994). To account for the possible strongly anisotropic scatterers, Hapke (2002) ultimately introduced an expansive treatment of multiple scattering (AMSA, or Anisotropic Multiple Scattering Approximation) that can be applied to strongly anisotropic grains.

---

[1] Muinonen et al. (2012) have recently derived a coherent-backscatter model on the basis of a rigorous solution of Maxwell's equation (see Section 2.3.3)

*2.3.2  Shkuratov Model*

The Shkuratov model (Shkuratov et al., 1999) also is physically motivated and describes a macroscopically fractal-like surface, and it includes a description of both SHOE and CBOE. Unlike the Hapke and Lumme-Bowell models in which the single-scattering and multiple-scattering contributions can be separated, the Shkuratov model is the product of three component functions and a normal reflectance coefficient $A_n$. In terms of the radiance factor

$$R(\mu_o, \mu, \alpha) = A_n f_{SHOE}(\alpha) \, f_{CBOE}(\alpha) \, d(\Lambda, \beta, \alpha) \qquad (14)$$

where the $f_{SHOE}$ describes the shadow-hiding opposition effect, $f_{CBOE}$ the coherent-backscatter opposition effect, and $d(\Lambda, \beta, \alpha)$ is the Akimov disk-distribution function that describes how the brightness of the surface at any given phase angle varies with photometric longitude, $\Lambda$, and photometric latitude, $\beta$. The Shkuratov model has four adjustable parameters. The coefficient $A_n$ is the normal albedo. The function $f_{SHOE}=\exp(-\kappa \, \alpha)$, where $\kappa$ is a parameter that decreases with increasing albedo to model the attenuation of projected shadows by multiple-scattering. The $f_{CBOE}$ function has two parameters: $\Lambda_E$, the extinction mean free path of a photon; and $d$, a size-scale parameter that defines the separation distance of scatterers that contribute to coherent backscatter interactions. The Akimov disk-distribution of brightness has no model parameters (Table 2).

*2.3.3  Muinonen RT-CB Model*

Many of the recent efforts to improve physically-motivated photometric models have focused on incorporating a description of the coherent backscatter opposition effect. In the analytical models above, the treatment of coherent backscatter is made tractable through the use of physically reasonable simplifying assumptions and mathematical approximations – none represent exact solutions of Maxwell's equations. At present, numerical methods provide the most rigorous approach to test and model how coherent backscatter operates in the context of planetary regoliths.

A breakthrough was achieved by Muinonen et al. (2012) who developed a Monte Carlo integration approach that combines radiative transfer and coherent backscatter (RT+CB) from an assumed spherical assemblage of randomly distributed spherical scatterers of known size, packing, and optical constants. A particularly important aspect of this approach is that it can be independently verified by direct computer solution of Maxwell's equations (Mackowski & Mishchenko 1996, 2011). While earlier works demonstrated the behavior of coherent backscatter among widely dispersed scatterers, the recent work of Muinonen et al. has definitively verified how the mechanism operates within loosely-packed aggregates of scatterers. Their Monte Carlo simulation accurately predicts not only the contribution of coherent backscatter to the photometric opposition effect, but also accurately models the polarization behavior at small phase angles (known as the polarization opposition effect) and the circular polarization. While most tests were done assuming uniformly-sized scatterers, in a preliminary experiment in which a size-distribution of scatterers was used, the results qualitatively matched the opposition behavior seen in high-albedo planetary objects, such as bright icy satellites.

The studies in Muinonen et al. (2012) were limited to sparsely-packed arrangements of scatterers (packing factors only as large as 6.25%). To be more directly applicable to regolith-covered bodies, more work will be needed to determine the method's applicability to densely packed arrangements of scatterers.

Further numerical modeling techniques were developed by Muinonen et al. (2011) and Wilkman et al. (2014a) with the lunar mare data acquired by SMART-1/AMIE. They employed a numerical approach to solve for the shadowing function resulted from the surface roughness using ray-tracking technique with the LS sphere as the auxiliary tool. Combining the numerical shadowing function solution and the RT-CB model with the classic 2PHG function and LS phase function as the volume phase function, Wilkman et al. (2014a) achieved good fit to the lunar mare data using Bayesian techniques and Markov chain Monte Carlo. Wilkman et al. (2014) provided a numerical representation of the shadowing function for close-packed spherical volume of LS particles to facilitate the photometric modeling of low-albedo surfaces.

2.3.4  IAU Magnitude Phase Curve System:

The Lumme-Bowell model (Lumme and Bowell, 1981b) is of special importance to asteroid studies because it was the original basis of the official IAU magnitude system for classifying and modeling the albedos and photometric behavior of whole-disk phase curves. Details of the Lumme-Bowell model and its adaptation to the H-G system are given in the *Asteroids II* volume of this series (Bowell et al., 1989). This long-used model was derived prior to the recognition of a narrow "opposition spike" that is caused by coherent backscatter and typically observed on airless bodies at phase angles less than a couple of degrees. Muinonen et al. (2010) developed a three-parameter model, called the H-$G_1$-$G_2$ model that is similar to the H-G model above but which significantly improves fits to asteroid phase curves that have a detectable opposition spike. In this model, that now officially replaces the previous IAU system, the reduced magnitude of an asteroid, $V(\alpha)$, is expressed as,

$$V(\alpha) = H - 2.5 \log_{10}[G_1 \Phi_1(\alpha) + G_2 \Phi_2(\alpha) + (1 - G_1 - G_2)\Phi_3(\alpha)] \quad (13)$$

where two numerically optimized, cubic-spline basis functions, $\Phi_1(\alpha)$ and $\Phi_2(\alpha)$ replace the basis functions of the H-G model, and a third basis function, $\Phi_3(\alpha)$, is introduced to describe the opposition effect. These basis functions are numerically defined by their values at their phase angle grids and the first derivatives at the two ends, as given in Tables 3 and 4 of Muinonen et al. (2010). The definition of *H* remains the same as in the H-G model. A few characteristic parameters of a phase function can now be expressed in terms of $G_1$ and $G_2$, including the phase integral, $q = 0.009082 + 0.4061 G_1 + 0.8092 G_2$, the photometric phase coefficient, $k = \frac{1}{5\pi} \frac{30 G_1 + 9 G_2}{G_1 + G_2}$, and the amplitude of the opposition effect, $\zeta - 1 =$. Since the $G_1$ and $G_2$ parameters of many asteroids appear to be somehow correlated, and can be described by a two-segment linear function (Muinonen et al., 2010; Oszkiewicz et al., 2011), the three-parameter model can be further reduced to a two-parameter model, parameterized by *H* and $G_{12}$. These new phase function models significantly improve the fits to existing asteroid phase function data over a wide range of phase angles up to 140º. It has been shown that asteroid taxonomy classes generally have their $G_1$, $G_2$ parameters, and therefore $G_{12}$ parameters as well, clustered in parameter space (Oszkiewicz et al., 2011; 2012). For example, the $G_{12}$ parameters of C-complex peak at ~0.65, while that of S-complex peak at ~0.45.

**2.4 Effects of Shape Models on Photometric Modeling**

Among the most important photometry-related innovations in recent years has been the development and standardization of three-dimensional digital shape models for asteroids, comets, and irregularly-shaped moons of the outer planets. With the advent of close-up imaging of asteroids and other irregularly-shaped bodies by spacecraft, radar observations and shape reconstruction from Earth, and sophisticated computer modeling of telescopic rotational lightcurves, it has become possible to measure the non-uniform topography of these objects and represent them in digital form. Details of these models and how they are derived can be found in the chapter by Ďurech et al. (this volume).

In photometric studies of asteroids from spacecraft missions, shape models play a critical role in the sampling of disk-resolved photometric observations from imaging data and spectrometers. That is, for disk-resolved observations, it is important to determine the photometric geometry (i.e. local angles of incidence, emission, and phase) for each point on the surface for which a corresponding brightness (i.e. radiance factor) is to be sampled. At any such point, the shape model, ephemeris and spacecraft navigation and pointing data provide the means to compute the orientation of the local vector surface normal from which photometric angles are measured (Fig. 1)

A pitfall in the use of shape models is that, at large incidence angles (to which properties like macroscopic roughness are very sensitive), even tiny errors in the estimated angle of incidence derived from the model can lead to large errors in photometric model fits. The presence of inaccurately represented topography, for example small craters or ridges that are not below the spatial resolution of the topographic model almost invariably introduce errors in the measured photometric angles. Consequently, it is common practice to exclude or otherwise apply weak statistical weighting to observations at large incidence angles from fits of photometric models to disk-resolved asteroid data.

**2.5 Testing of Physically-Motivated Models**

*2.5.1 Hapke Model Controversy*

Shkuratov *et al.* (2012, 2013; see also Zhang and Voss 2011) criticized Hapke's model on the basis of 15 alleged failures and deficiencies. While it is beyond the scope of this chapter to provide a point-by-point discussion, these criticisms fall into four categories; that Hapke's model 1) is not rigorous in its application of electromagnetic theory and it violates conservation of energy, 2) it makes unrealistic assumptions about light scattering by natural, particulate regolith surfaces and it incompletely describes the dependence of scattering on photometric geometry, 3) it employs too many adjustable parameters, some of which are not truly independent variables, making it cumbersome and ambiguous to use, and 4) in the authors' tests the model exhibits unrealistic behavior in comparison to their ray-tracing predictions, and show that parameter values retrieved from fits of the model to the laboratory and lunar photometric data are unreliable.

Hapke (2013) argued that nine of the 15 criticisms are invalid because they are based on a profound misunderstanding of his model and the assumptions that he made in deriving its components. He acknowledged that the greatest weakness in his model is that the coherent backscatter opposition effect as it applies to regolith surfaces is poorly understood and consequently its treatment in his model is imperfect and necessarily preliminary. He notes that CBOE is known to occur in regolith surfaces and it was necessary to include at least a preliminary description to model observed regolith opposition effects. Hapke does not dispute that his model is not rigorously derivable from Maxwell's equations. He argues, though, that his model is nonetheless useful when it is applied and interpreted within the limitations that are imposed by his simplifying assumptions, all of which are stated in his published derivations. In response to one criticism, Hapke acknowledged that his recent porosity correction indeed violates the conservation of energy at distances that are smaller than the inter-particle separation distance. However, he points out that the violation is irrelevant because the definitions in his model are specifically made such that energy is conserved over all distances larger than the inter-particle separation, which is the minimum distance over which the radiative transfer equation is applicable.

In practice, the most difficult criticism for users of the Hapke model to overcome is that its numerous coupled (i.e. non-independent) model parameters can lead to ambiguous or unreliable parameter value retrievals when applied to observations of planetary regoliths (see below). Hapke attributes the mutual dependence of the modeled regolith properties to a true property of nature and not a failure of his model. The ability to reliably retrieve meaningful values of model parameters first and necessarily depends on the availability of sufficient photometric data coverage to fully constrain the model parameters – coverage that is often neither fully available from photometric observations of planetary surfaces, nor is the adequacy of the available data to constrain the Hapke model fits often tested by practitioners before application. Secondly, judicially chosen fitting methods can often overcome difficulties with potentially ambiguous model retrievals (cf. Hapke 2012a, Hapke et al. 2012). In addition it is sometimes possible to constrain some of the model parameters, even when data are insufficient to simultaneously constrain all of them by exploiting ranges of photometric angles for which the unconstrained parameters have insignificant effect (cf. Souchon *et al*. 2011). Finally, as discussed in Helfenstein and Shepard (2011), it is critically important to perform a thorough error analysis on the results so that ambiguous or unreliable values can be identified.

*2.5.2   Recent Tests*

Investigations to validate the fidelity and physical interpretability of physically-motivated photometric models fall into three broad categories; laboratory studies, computer modeling studies, and *in situ* studies.  The focus of many laboratory studies has been to test and validate the Hapke model (Buratti and Veverka, 1985; Kamei and Nakamura 2002; Cord et al., 2003; 2005; Gunderson et al., 2005; Shkuratov et al., 2007; Shepard and Helfenstein, 2007; 2011; Helfenstein and Shepard, 2011;  Hapke et al., 2009; Souchon et al., 2011) as well as studies that compare the predictions of Hapke's model to numerical light-scattering simulations (Helfenstein, 1988; Shepard and Campbell, 1998; Ciarniello et al., 2014). Finally *in situ* studies rely on using well-sampled planetary observation data sets to derive information about the validity of Hapke's model (cf. Guiness et al., 1997; Verbiscer and Veverka, 1990; Domingue et al., 1997; Helfenstein et al., 1997; Helfenstein and Shepard, 1999; Hapke et al., 2012; Shepard et al., 2001; 1993).

Studies that exploit *a priori* knowledge of the optical properties of soil components to perform forward-modeling of surface reflectance are often (but not always) more successful than efforts that aim to invert the Hapke model to assess properties of particulate surfaces from photometric observations. Forward-modeling of remotely-sensed reflectance spectra with Hapke theory has been routinely accomplished for many years for a wide range of planetary objects (cf. Roush et al., 1990; Calvin and Clark, 1991; Lucey, 1998; Verbiscer et al., 2006; Wilcox et al., 2006; Denevi et al., 2008; Warell and Davidsson, 2010). The model successfully predicts the scattering behavior of idealized soil samples (Hapke et al., 2009; Hapke and Wells, 1981; Hapke, 2012b). Computer simulation studies have investigated the geological interpretation of Hapke's (1984) correction for macroscopic roughness (Helfenstein 1988; Shepard and Campbell 1998). More recently, for lunar soils and laboratory analog materials, photometric roughness best matches lunar regolith relief at submillimeter size-scales (Helfenstein and Shepard, 1999; Cord et al., 2003; Goguen et al., 2010). Buratti and Veverka (1985) showed that photometrically detected roughness is sensitive to the albedo of the surface, with high albedos causing dilution of shadows via multiple-scattering. Tests of the opposition surge model conclude that both coherent backscatter and shadow hiding contribute to the opposition effect (Helfenstein et al., 1997; Hapke, 1998; Hapke et al., 2012). Hapke et al. (2012) found that the angular width of the lunar opposition effect does not significantly vary with optical wavelength, contrary to the current analytical treatments of coherent backscatter, which, for the Moon, predict that the width should be proportional to the square of the wavelength. In combination with experimental evidence, the observed absence of a wavelength-dependence to the opposition effect suggests that our current understanding of coherent backscatter is deficient. In Monte Carlo simulation experiments verified by direct solution of Maxwell's equations, Muinonen et al. (2012) modeled scattering from a spherical volume of sparsely distributed smaller spherical scatterers (see Section 2.3.4). They found that the angular width of the CBOE was independent of the packing density of small scatterers, which is best explained if the mean length of the interference base is controlled by the size-parameter of the spherical scattering volume rather than the transport mean free path that is often assumed in analytical approximations.

Attempts to validate Hapke's (1981; 1984; 1986; 2002) theory by comparing the known characteristics of laboratory samples to those predicted by fits of the model parameters to photometric observations of the samples found no compelling evidence that individual photometric parameters could be uniquely interpreted to reveal the physical state of the samples, either in an absolute or relative sense. Instead, combinations of physical properties such as particle single-scattering albedo, roughness, and porosity were convolved within each retrieved photometric parameter (Gunderson et al., 2005; Shepard and Helfenstein, 2007; Shkuratov et al., 2007, 2012; Souchon et al., 2011). In general, the best agreement was found, at least qualitatively between the observed microstructure of natural soil grains and the physical interpretation of retrieved particle phase function model parameters (Souchon et al., 2011; Shepard and Helfenstein, 2007).

Only recently have tests of Hapke's (2008) correction for the effects of porosity been published (see also Section 2.5.1). Helfenstein and Shepard (2011) performed a preliminary test of the model by applying it to analyze laboratory analog samples from their 2007 suite of measurements (Shepard and Helfenstein, 2007). Their results suggest that Hapke's porosity correction improves the fidelity of fits to samples composed of low- and moderate-albedo particles and may allow for more reliable retrieval of porosity estimates in these materials. However, the test also suggested that in high-albedo surfaces, the effects of porosity may be difficult to detect. More recently, Ciarniello et al. (2014) use Monte Carlo simulations of ray-tracing in particulate media to test three formulations of the Hapke's model: the multiple-scattering IMSA, AMSA, and an updated version of Hapke (2008). The found that, excluding the opposition effect, the Hapke (2008) model is the most accurate to describe the reflectance behaviors of particulate media with arbitrary porosities. It is also able to characterize anisotropic scattering, unless the medium exhibits a strongly forward-scattering behavior. The IMSA and AMSA models were effective only for soils with extremely large porosities.

### 3. OBSERVATIONAL DEVELOPMENTS

Prior to the advent of interplanetary space probes, all photometric observations of asteroids were telescopic in which objects were seen only as point sources. Thus, only disk-integrated data were available. Except for Earth-crossing asteroids, the observable phase angle range was generally limited to less than 25°-30°. When spacecraft encounters with asteroids became reality, it was possible to measure an asteroid's three-dimensional shape, to observe how the brightness of the asteroid's surface varied with local photometric geometry, and to detect terrain-dependent albedo and color heterogeneities.

Spacecraft data have their own limitations in photometric modeling, too. For some orbit missions, such as Dawn (Russell et al., 2012; also in this volume) that use polar orbits, there may be a correlation between latitude and phase angle since the cameras observe the surface in the nadir direction ($e \approx 0°$) (e.g., Li et al., 2013a). When the field-of-view of the camera is small compared to the angular size of the target, this can introduce a bias to the observations. For fast flyby observations, the illumination geometry does not change much from image to image taken within a short duration (typically <1 hour), which is much shorter than the rotation period of most asteroids, and the flyby. Derivation of a shape model may then be restricted by very limited illumination and surface coverage, and the photometric analysis would be subject to the same limitations. Spatial resolution and uncertainties in shape can also limit the accuracy of photometric models. These limitations should be considered in the modeling and interpretation of disk-resolved photometric data.

Since the first asteroid flyby of (951) Gaspra by the Galileo spacecraft in 1991 (Belton et al., 1992; Veverka et al., 1994), thirteen asteroids have been imaged by spacecraft either from flybys or rendezvous. Seven asteroids were visited since the publication of the *Asteroids III* book in this series, and Ceres will be visited by Dawn starting in early 2015. Of the seven newly imaged asteroids, the data of (25143) Itokawa (Yoshikawa et al., this volume; Kitazato et al., 2008), (2867) Šteins (Barucci et al., this volume; Spjuth et al., 2012), (21) Lutetia (Barucci et al., this volume; Magrin et al., 2012, Masoumzadeh et al., 2014), and (4) Vesta (Russell et al., this volume; Li et al., 2013a; Schröder et al., 2013; Longobardo et al., 2014) are of sufficient resolution and quality to perform detailed, disk-resolved photometric studies. The images of (5535) Annefrank from Stardust flyby were of low resolution, and only disk-integrated analysis has been performed from the spacecraft data (Hillier et al., 2011). The flyby of (132524) APL by New Horizons was serendipitous at a large distance and the returned images resemble point sources, and therefore no photometric modeling was performed. No report about the photometric properties of (4179) Toutatis from the Chinese Chang'E 2 flyby images (Huang et al., 2013; Barucci et al., this volume) has yet become available. (1) Ceres was observed by HST in 2003/04 with the Advanced Camera for Surveys (ACS) High Resolution Channel (HRC), and resolved to ~30 km/pix, facilitating the study of its disk-resolved photometric properties (Li et al., 2006). In this section, we will summarize the photometric results for Ceres, Itokawa, Šteins, Lutetia, Vesta, and Annefrank.

**3.1 Ceres**

Ceres represents a thus-far unique category of asteroids or dwarf planets (cf. Rivkin et al., 2011). Ample evidence suggests that water must have played an important role in its evolutionary history and its current status (Rivkin et al., 2011). The recent unambiguous discovery of water vapor associated with localized sources on Ceres indicates a possibly active surface at the present (Küppers et al., 2014).

Li et al. (2006) reported a detailed photometric analysis of Ceres based on the images collected by the HST/ACS through three broadband filters centered at 535, 335, and 223 nm obtained in 2003/04 at a narrow range of phase angles of 5°-7°. The disk-resolved analysis focused on studying the limb-darkening behavior to derive the roughness, and determining the single-scattering albedo. Li et al. reported a Hapke roughness parameter of 44°±5' for Ceres, much higher than those of all other objects that have been modeled with Hapke functions that include roughness (Table 6). Given that the roughness of all other objects were derived from images with much higher spatial resolution than those of Ceres, it is a question whether this high roughness is real or an observational artifact. Indeed, the small angular size of Ceres (~30 pixels across) and the point spread function of the optical system could blur the edge of Ceres' disk, so that the brightness decreases towards the limb is much faster than the real surface, yielding higher roughness in the modeling. This effect probably also explains the fact that the Hapke and Minnaert model could only fit Ceres' disk within 40°-50° photometric longitude/latitude as shown by Li et al. On the other hand, the high photometric roughness of Ceres is consistent with its high radar roughness of 20°-50° (Mitchell et al., 1996), although the same radar observations also indicate that the surface of Ceres is very smooth at centimeter to decimeter scales, probably indicating different roughness at different scale sizes.

The photometric analysis in the visible and near-infrared wavelengths suggested a remarkably uniform surface of Ceres in its albedo and color (Li et al., 2006; Carry et al, 2008; 2012). The albedo maps at all three wavelengths show variations of only 4% full-width-at-half-maximum (FWHM), and about ±6% peak-to-peak. Compared with other solar system objects with their surface photometric variations determined, Ceres certainly has the most uniform surface so far (Li et al., 2006). Of course, with much higher spatial resolutions in the data to be collected by Dawn, it cannot be excluded that some local areas on Ceres that are much smaller than the HST image resolution have larger albedo and/or color variations from the global average. But those small areas should not dramatically change the overall distribution of the surface albedo and color on Ceres.

### 3.2 Itokawa

Lederer et al. (2005) used broadband BVRI photometry of the S-type asteroid Itokawa between 4º and 90º phase angles obtained from the ground to characterize this target asteroid of Hayabusa prior to its launch. The authors used a Hapke model to derive a single scattering albedo of 0.36, and a highly backscattering single particle phase function with $g$=-0.51, although the roughness and opposition surge parameters were assumed and could have driven this number to be large. Lederer et al. found that Itokawa has a higher albedo than average main-belt S-class asteroids. However, limited by disk-integrated observations and the lack of data at small phase angles, their models have to assume an opposition surge and the common values of roughness of 20º and 36º.

The Japanese Hayabusa spacecraft visited Itokawa in 2005 (Yoshikawa et al., this volume; Fujiwara et al. 2006). Kitazato et al. (2008) conducted a photometric analysis of Itokawa, using disk-resolved reflectance spectra measured by the Near-infrared Spectrometer (NIRS) over multiple wavelengths ranging from 0.85 to 2.10 μm covering phase angles from near 0º to 38º. Using a Hapke photometric model analysis, the authors found that Itokawa has a single-scattering albedo that is 35-40% less than that of asteroid 433 Eros, consistent with the results of Lederer et al. (2005). The data at low phase angles enables modeling of opposition effect. While Kitazato et al. only included SHOE but not CBOE in the Hapke modeling, they observed wavelength dependence of the width of opposition effect, and attributed it to coherent backscattering. Compared to Eros, the lower opposition amplitude indicates weaker contribution of the SHOE component, and the narrower width is consistent with relatively more densely packed particles in the regolith of Itokawa. The roughness parameter for Itokawa was estimated to be 26º, similar to that of Eros'. The modeled parameters of opposition effect and roughness justify the assumptions made in Lederer et al. (2005) work, although those studies were performed at a different wavelength regime.

### 3.3 Šteins

Spjuth et al. (2012) performed detailed photometric analyses of asteroid Šteins, the first E-type asteroid visited by spacecraft, using multi-spectral images acquired by the OSIRIS Wide Angle Camera (WAC) onboard the Rosetta spacecraft (Keller et al., 2007). An extremely wide range of phase angles from 0.36º to 130º was achieved through one filter centered at 630 nm, allowing for modeling of the opposition effect. The authors estimated the single scattering albedo to be 0.57, the highest ever found for small bodies visited by spacecraft. The roughness is 28º, and the asymmetry factor of the 1PHG single-particle phase function is -0.27. The geometric and Bond albedos of Štein were derived to be 0.39± 0.02 and 0.24±0.01, respectively. The high albedo of Šteins is consistent with the iron-poor surface composition of the asteroid (Barucci et al. 2012). To study the photometric variations across the surface of Šteins, Spjuth et al. generated maps of Hapke photometric parameters, including the single-scattering albedo, roughness, and the asymmetry factor of the 1PHG function, by assuming that the reflectance variation is entirely caused by one single parameter while keeping the others fixed at their respective global averages. These maps probably represent the highest possible variations in each single photometric parameter. As indicated by these maps, the surface of Šteins seems to be highly homogeneous, with a standard deviation of 0.01 for its single-scattering albedo, 3.4° for roughness, and 0.003 for the asymmetry factor. The uniform surface of Šteins is consistent with the results reported by Leyrat et al. (2010), who applied photometric correction to the flyby images with a Hapke (2002) model and performed a statistical analysis, and find no surface inhomogeneities larger than 4% at the 95% confidence level.

The photometric properties of Šteins are interpreted as the consequence of its regolith properties (Spjuth et al., 2012). The strong opposition effect and high albedo are consistent with porous, fine-grained regolith, which is consistent with the studies of aubrite meteorites, which are considered to be analog to E-type asteroids, and laboratory work (Gaffey et al., 1989; Shepard and Helfenstein, 2007). The low amplitude of SHOE indicates relatively transparent regolith particles, and/or a large density of internal scatterers of opaques. The width of opposition effect is also consistent with a high porosity of ~84% for the regolith particles.

**3.4 Lutetia**

Belskaya et al. (2010) obtained ground-based BVRI photometric and V-band polarimetric measurements of Lutetia over a wide range of phase angles (0.5°-22.2°). The authors found that Lutetia has a non-convex shape and heterogeneous surface properties, due possibly to a large crater and the variations in the texture and/or mineralogy, respectively. The polarimetric properties of Lutetia suggest a fine-grained regolith with a mean grain size smaller than 20 μm, and of carbonaceous composition (CO, CV, CH). The Rosetta spacecraft flew by Lutetia at a close encounter distance of ~3200 km on July 10, 2010. The flyby geometry at Lutetia is similar to that at Šteins, while multicolor images were acquired through both WAC and the narrow angle camera (NAC) throughout the flyby, providing phase angle coverage from less than 1° to 130° at multiple wavelengths. Based on these data, La Forgia (2014) conducted a photometric analysis of Lutetia with a disk-integrated Hapke model. They showed that all the modeled parameters depend on wavelength. However, it has to be noted that their modeling is based on disk-integrated phase functions, and therefore strong coupling exists between at least the roughness parameter and the phase function. The photometric analysis of Lutetia showed that the asteroid's surface is composed of particles that have high reflectivity. La Forgia (2014) also found that Lutetia's surface is generally smooth and characterized by a low porosity. These results are consistent with the conclusion that Lutetia is an ancient planetesimal rather than a reaccumulated object, as indicated by its large diameter of 98±2 km (Sierks et al. 2011a) and its high density of 3.4±0.3×10$^3$ kgm$^{-3}$ (Pätzold et al. 2011). Masoumzadeh et al. (2014) performed a disk-resolved photometric analysis with the Hapke model and the Minnaert model using the Rosetta flyby data. Different from the La Forgia (2014) results, they reported no wavelength dependence on the Minnaert *k* parameters. In addition, in their Hapke modeling, they derived very different single-scattering albedo with 1PHG and 2PHG single-particle phase function (0.26 vs. 0.49, respectively).

### 3.5 Vesta

Dawn's rendezvous with Vesta provided us with high-resolution imaging data (Russell et al., 2012; 2013; also see the Dawn chapter in this volume). Li et al. (2013a) performed a thorough analysis of the global photometric properties of Vesta using all Dawn Framing Camera (FC; Sierks et al., 2011b) images with pixel scale >0.25 km/pixel through all filters, including one clear filter and seven mid-band (~100 nm bandpass) color filters with center wavelengths from 440 nm to 960 nm. Dawn data cover a range of phase angles from ~7° to ~90° for disk-resolved data in all filters, and from ~24° to ~108° for disk-resolved data through the clear filter. Li et al. found that the photometric behavior of the surface of Vesta is well described by a Hapke model and a Minnaert model, but the LS model cannot reproduce the limb-to-terminator brightness trends. The phase functions of Vesta show weak dependence on wavelength, qualitatively consistent with previous ground-based observations (Reddy et al., 2012a), but about a factor of 0.5 weaker (Li et al., 2013a). The phase reddening of Vesta as observed by Dawn is similar to, or slightly stronger than that of Eros observed by the NEAR mission (Clark et al., 2002). The Hapke roughness parameter of Vesta is consistent at all wavelengths studied. While roughness parameter is a geometric parameter and is not supposed to depend on wavelength, for high albedo surfaces, multiple scattering into shadows could in principle wash out shadows and result in lower modeled roughness. The lack of such effects on Vesta suggests that this effect is small for an SSA of <0.55 and a roughness of ~20°. Compared with other asteroids studied so far, the globally averaged photometric properties of Vesta are similar to those of S-type asteroids, except for its albedo, which is about twice the average albedo of S-type's.

Vesta has a highly heterogeneous surface in almost all aspects (Reddy et al., 2012b; Jaumann et al., 2012; Prettyman et al., 2012; De Sanctis et al., 2012a; 2013; Ammanito et al., 2013). Its albedo and color variations have been recognized previously from ground-based and HST observations (Gaffey, 1997; Binzel et al., 1997; Li et al., 2010; Reddy et al., 2013). The albedo and color maps generated from Dawn FC data further show that such heterogeneity is at almost all scale sizes, and reveal many localized areas with distinctly high and low albedos (McCord et al., 2012). The overall distribution of albedo on Vesta has a FWHM of ~17%, but the localized bright and dark areas have albedos from about a half to greater than twice the average (Li et al., 2013a). The single-peaked albedo distribution on Vesta is presumably a result of strong regolith mixing on Vesta caused by the complicated impact history of Vesta (Pieters et al., 2012), different from the bimodal distribution on the Moon as dominated by compositional heterogeneity (e.g., Helfenstein and Veverka, 1987; Hillier et al., 1999; Yokota et al., 2011).

Schröder et al. (2013) reported correlations between the spatial variations in the surface phase function of Vesta and geology. After correcting the limb-to-terminator brightness trend on Vesta, they were able to map out the slopes of the surface phase functions over the whole illuminated surface of Vesta. The slope of phase function shows clear global distribution correlated with albedo and mineralogical units. Eucrite-dominated regions (Ammanito et al., 2013) have relatively steeper phase functions. Since those regions are also correlated with the enriched distribution of dark material, which is interpreted as exogenous carbonaceous material delivered by impactors (McCord et al., 2012; Reddy et al., 2012c; Prettyman et al., 2012; De Sanctis et al., 2012b; Turrini et al., 2014), such correlation between albedo and phase function on Vesta probably manifests the general correlation in the asteroid belt (e.g. Bowell and Lumme, 1979).  Furthermore, in local scales, the phase function slopes are steeper on crater floors than on crater walls, interpreted as a variation in roughness due to the accumulation of loose material inside craters causing higher roughness.

Longobardo et al. (2014) and Li et al. (2013c) analyzed the visible and NIR data collected by Dawn's Visible and Infrared (VIR) instrument (De Sanctis et al., 2011), but this work is still ongoing as of the writing of this chapter.  The preliminary results reported by Li et al. (2013c) using VIR data were inconsistent in the overlapping wavelengths with those derived from the FC data taken at similar mission phases, tentatively attributed to instrumental effects. Longobardo et al. (2014) took a completely empirical approach with Akimov model to fit VIR data and study the dependence of pyroxene band parameters on scattering geometry with the focus on bright and dark regions on Vesta.  Their results should be less dependent on instrumental effects.  The trend of band parameters with respect to scattering geometry derived from VIR data is shown to be consistent with previous observations from the ground (Reddy et al., 2012a).

### 3.6 Annefrank

Disk-integrated Stardust observations of Annefrank obtained between 47° and 135° were combined with ground-based measurements obtained between 2° and 18° and analyzed by Hillier et al. (2011).  This asteroid exhibits a steep phase curve, resulting in a very large roughness parameter of 49°, an unusually large single particle scattering albedo of 0.62, and a Henyey-Greenstein phase function, $g$, of -0.09, which is more isotropic than the usual strongly backscattering S-type asteroids.  The shadow-hiding opposition surge width $h$=0.015 is more typical of the S-types.  Because the shape of Annefrank is irregular, these fits to a full excursion in solar phase angle may not be unique, especially with disk-integrated data only.  Li et al. (2004) have shown that the assumption of sphericity can lead to an overestimate of both the single scattering albedo and the roughness parameter and a more isotropic phase function, particularly if measurements at large solar phase angles are included in modeling.  Hillier et al. restricted their fits to 90° and found more typical parameters of 20° for the roughness parameter, 0.41 for the single scattering albedo and -0.19 for $g$.  Still, the single scattering albedo of Annefrank is high.

### 4. APPLICATIONS AND INTERPRETATIONS

### 4.1 Correction for spectral analysis

*4.1.1 Photometric correction*

Most of the changes in the reflectance of an asteroid surface are not intrinsic but rather due to variations in viewing geometry.  With a description of the photometric functions of the surface, all the non-intrinsic variations can be removed and a map of the normal reflectance or a physical parameter such as the single scattering albedo can be produced.  Since the three geometric parameters for the scattering, $i$, $e$, and $\alpha$, are known *a priori* (if one possesses an accurate shape model for non-spherical bodies, which is often a requirement for asteroids), fitting a photometric function or model to spacecraft or ground-based images provides all the information to perform photometric corrections.

The usual procedure for photometric correction as in Eq. 1 is: 1) Use the observed data to fit a global average photometric model; 2) Calculate the ratio of the measured reflectance over the model reflectance for each surface element; and 3) Multiply the ratio by the model reflectance at the desired reference geometry. In this procedure, the ratio calculated is a measure of the deviation between the actual surface reflectance and the best-fit model, and the user has a choice of reference geometries for the correction. The implicit assumptions behind the photometric correction are: 1) Any deviation between the measured reflectance and the best-fit model must be real, and not due to imperfections in the photometric model or the shape model; and 2) The deviation factor is independent of scattering geometry. We shall now discuss the implications of these two assumptions.

The first assumption can sometimes be verified by fitting several different models to the same data and comparing the resultant photometric corrections, but this is often limited by data availability. Therefore, one has to judge what features in the photometrically corrected maps are real and what are artifacts based on the magnitude, scale size and shape, the geologic and compositional context, etc., on a case-by-case basis. In our discussion about the second assumption (next), we will assume that the first assumption is true.

For the second assumption, two different cases exist. The first case is that the reflectance variation from the best-fit model prediction is dominated by the single-scattering albedo variation rather than variations in other photometric properties such as phase function or roughness. If albedo is low and multiple scattering is insignificant, then the reflectance is approximately proportional to the single-scattering albedo (see Section 2). Thus, the deviation factor represents how much the albedo at a particular surface location deviates from the global average, and should not depend on scattering geometry. The assumption is justified. On the other hand, for high albedo cases where multiple scattering is significant, the proportionality between single-scattering albedo and reflectance is broken, and photometric correction needs to be performed taking into account the non-linear effect. This is a complicated process but feasible. The second case is the opposite of the first case: the variation in phase function and/or roughness dominates the deviation factors. Since the effects of phase function and roughness always depend on phase angle, so does the deviation factor. In this case, photometric correction cannot be reliably applied. For example, it is well known that variations in photometric properties other than albedo exist on the Moon, where multiple photometric models are derived for different regions based on their albedos, and photometric corrections are performed based on the corresponding models (e.g., Yokota et al., 2011; Besse et al., 2013; Wu et al., 2013). Schröder et al. (2013) work also shows that this case exists on Vesta. One has to be aware of this constraint when interpreting data products that involve a photometric correction of a bright surface.

The selection of an appropriate photometric model for the surface is important when performing a photometric correction. In principle, any photometric model that can describe the photometric behavior of the target asteroid surface sufficiently well can be used for photometric correction, whether or not physical interpretations of model parameters are possible. Although radiative transfer models can be used to derive intrinsic physical parameters such as the single scattering albedo or normal reflectance, unless a wide range in solar phase angles is available, unique fits to parameters in Hapke-type models are not possible (Helfenstein et al., 1988). For practical photometry, empirical functions are often preferred due to their simple mathematical forms and fewer free parameters.

*4.1.2 Phase reddening*

It is well known that spectra of asteroids, and many other airless bodies, become redder as the solar phase angle increases (see review in Gehrels, 1970; Gradie and Veverka, 1986). This phenomenon of "phase reddening" is most pronounced in S-type asteroids: Bowell and Lumme (1979) found that between 0.35 $\mu$m and 0.55 $\mu$m the effect was 2-3 times greater for S-type asteroids than for C- or M-types. It is generally considered that phase reddening is related to the increase in spectral reflectance with wavelength. For example, the albedo of S-type asteroids doubles between 0.35 $\mu$m 0.8 $\mu$m (Tedesco et al., 1989). The physical reason for the reddening probably lies in the relative importance of multiple scattering: as the albedo increases, the amount of radiation that has been multiply scattered also increases. The singly scattered radiation is backscattering, while the multiply scattered radiation tends to be isotropic. Thus, at large solar phase angles, in the forward-scattered direction, the multiply scattered photons are relatively more important. Since there are more such photons as the albedo increases, the reflectance becomes redder at large phase angles. The spectra of S-type and A-type asteroids exhibit the most significant changes in albedo through the visible range of the spectrum, so they should exhibit more phase reddening than the flat C- or M-types. The spectrally unique C-class asteroid Ceres has an almost featureless spectrum between 0.4 $\mu$m and 2.4 $\mu$m and it exhibits no phase reddening (Tedesco et al., 1983). But the connection between phase reddening and albedo is not straightforward. For example, the bright V-type asteroids, for which a composite solar phase curve was constructed from NEOs, exhibit phase reddening of only 0.002 mag/deg in B-R and 0.0016 mag/deg in I-R, even though this class of asteroids typically changes in albedo by a factor of 2 between the B and I filters (Hicks et al., 2014).

Even with a spectrally flat body it is possible to get phase reddening. At larger solar phase angles, shadows cast by rough features occupy an increasing fraction of the face of the object as seen by an observer. Although these shadows are dark shadows for low-albedo surfaces, they are partly illuminated by multiply scattered photons for high-albedo objects. Thus, only low-albedo objects with flat spectra should exhibit no solar phase reddening, which is the case for Ceres, which has a visible geometric albedo of 0.09.

**4.2 Taxonomic classes**

The application of photometric models to spacecraft images, coupled with the acquisition of measurements of Near Earth Asteroids (NEAs) at large solar phase angles not attainable from the ground for asteroids in the Main Belt, has led to the robust modeling of a sufficient number of asteroids so that trends among the various classes of asteroids can be studied. Asteroid spacecraft observations that have been fit with photometric models are listed in Table 6, together with the modeling results. Photometric modeling of ground-based measurements of NEAs has been done for (1862) Apollo, a rare Q-type (Helfenstein and Veverka, 1989). In addition, a composite solar phase curve of V-type asteroids was created from NEA observations and fit with both Hapke and Lumme-Bowell models (Hicks et al, 2014). The phase functions of C-, S-, and V-type asteroids are plotted in Figure 2.

[Table 6]

Asteroid (1862) Apollo, an NEA that is a member of the rare Q-type class, was observed extensively during its apparitions in 1980 and 1982 (Harris et al., 1987). Q-types are similar to S-types, but have higher albedos and are consistent with un-space-weathered materials, relative to the S-types (Binzel et al., 2010). The Q-types have a slightly more isotropic phase function, due possibly to this higher albedo: multiply scattered photons tend to be more isotropic than singly scattered photons. The smaller roughness parameter may also be the result of the asteroid's higher albedo, as multiply scattered photons partly illuminate shadows cast by rough features and lead to an underestimate of the roughness parameter.

Helfenstein and Veverka (1989) derived an average S and C type asteroid over the range of solar phase angles for the main belt (~0-30°). The small number of S-type asteroids that have been observed with spacecraft, and thus over a larger excursion in solar phase angles, have not shown a strong correlation in derived Hapke parameters (Table 6). This lack of correlation may be due to different solar phase angle sampling, or to non-unique fits, or to the models not fully accounting for the effects of multiple scattering, which partly illuminate primary shadows for both the roughness model and for the opposition surge shadow-hiding model. Or it is due to the lack of correlations between Hapke parameters and physical properties (Shepard and Helfenstein, 2007). It is also important to separate NEA and Main Belt populations because the NEAs are not as weathered and are smaller and thus expected to have less substantial regoliths.

Nevertheless, Belskaya and Shevchenko (2000) noted the correlations in Figure 3 of the Lumme-Bowell two-parameter IAU convention. Higher albedo asteroids have larger single scattering albedos, and larger slope parameters (G-parameter). The Henyey-Greenstein parameter is slightly more forward scattering for high-albedo asteroids, which is to be expected if they are more isotropic. The opposition surge parameter also tends to decrease as the albedo increases, with a slight decrease in the compaction state. The only C-type asteroid modeled (Mathilde) seems to confirm this trend: its opposition surge amplitude of 3.18±1.0 (Clark et al. 1999) is significantly higher than the typical S-type of 1-2. Whether this effect is due to actual physical changes or a decrease in multiple scattering, which is not fully accounted for in Hapke's models, remains to be seen as more asteroids are studied using radiative transfer models over more uniform phase angle sampling. Ceres, an atypical asteroid that is difficult to classify but is akin to the C-types, has an amplitude of only 1.6, so perhaps as more observations are gathered, Mathilde will be shown to be anomalous.

Benner et al. (2008) found a correlation between the roughness of NEA asteroid classes at radar wavelengths and asteroid class, with E and V-types being the roughest. However, in visible wavelength imaging data, Vesta appears to have a typical (not unusually rough) surface (Li et al, 2013a). This could be explained by the infilling of facets and asperities on the surface by fine particles, similar to the "ponding" seen on Itokawa and Eros, but on a smaller scale, so as not to be detected by radar wavelengths. With the larger gravity field of Vesta, small particles would be more likely to be retained than on the smaller asteroids.

## 4.3 Context of geological process

### *4.3.1 Roughness*

The roughness and size scale of features on planetary surfaces yield clues to a wide range of geophysical processes. Heavily cratered surfaces tend to exhibit large slope angles, while bodies that have undergone resurfacing due to volcanism are smoother. The accretion of dust or volatiles onto a surface could also infill facets to form a smooth surface. For example, the low-albedo hemisphere of Iapetus is smooth because features have been infilled with dust from Saturn's Phoebe ring (Lee et al., 2009). Regions with low surface roughness could be indicative of recent activity involving possible debris flows or ponding (Veverka et al., 2006; Robinson et al., 2001). On Vesta, crater walls tend to be smoother than crater floors due to the accumulation of mass-wasted materials in the crater (Schröder et al., 2013). Differences in roughness also provide a basis for chronology of various terrains, as rougher surfaces have been subjected to more impacts. Global roughness is also affected by the mass of the body: stronger gravitational wells tend to hold particles that are the result of meteoritic bombardment and that infill surface asperities. Viscous relaxation, which is a function of surface temperature and composition, can also smooth out a surface.

### *4.3.2 Particle size and packing status*

Irvine's (1966) model of the opposition shadow hiding effect assumes that particles comprising the regolith are in the geometric optics limit ($r>>\lambda$), spherical, single-sized, and sufficiently low in albedo so that multiple scattering can be neglected. The amplitude of the surge is determined primarily by a compaction parameter, defined by $(3/4\pi)(\varrho/\varrho_0)$, where $\varrho$ is the bulk density of the regolith particles and $\varrho_0$ is the density of an individual particle. Irvine's model is an analytical one that keeps track of shadows cast by regolith particles, as well as the fraction of individual particles that are visible to the observer and illuminated, as a function of solar phase angle. As compaction parameter increases and the surface becomes less porous, the amplitude of the opposition surge decreases. For Hapke's model of shadow hiding the amplitude of the opposition effect, $B_{0,SH}$ depends on the transparency of regolith grains; $B_{0,SH}=0$ for ideally transparent grains, and $B_{0,SH}=1$ for opaque ones. In a collection of uniform particles with no multiple scattering, the angular width parameter $h_{SH}$ is equal to $(3/8)K(1-\varrho/\varrho_0)$ (Hapke, 2012b). As the voids increase, $h_{SH}$ decreases. In addition, Hapke's latest model (2008; 2012a) presents a more complicated picture where, in general, increasing filling factor (decreasing the porosity) increases the reflectance of low and medium albedo powders, but decreases reflectance for ones with very high albedos.

Geophysical processes that cause porous surfaces include micrometeoritic bombardment, fallout of frost or dust from volcanic processes, and infall of ambient dust from interplanetary space or from rings. Mechanisms causing more compacted surfaces include annealing by high-energy particles, extrusion of fluids or dust-laden slurries, and viscous slumping. Intermediate porosities are caused by processes such as fluid-like flow of dust and "ponding" of regolith particles.

In the geometric optics limit it is difficult to derive particle sizes, not only because the shadowing process is scale invariant, but also because different combinations of size-distributions give similar porosities. The current shadow-hiding models also do not fully account for the effect of multiply scattered photons. These photons will partly illuminate primary shadows so that additional photons are scattered back into the direction of the observer with the result that the porosity will be underestimated.

## 5. SUMMARY AND FUTURE PERSPECTIVE

In conclusion, we have summarized the most recent developments in asteroid photometry since the publication of the *Asteroids III* volume. Advancements have been made in all three areas of theory, laboratory measurement, and spacecraft observation.

We have shown that photometric modeling is important for understanding what the light scattering behavior of an asteroid surface can tell us about the physical and compositional properties of the asteroid. We have shown that photometric modeling is important for "correcting" all spectral and imaging data to a common reference viewing geometry before inter-comparisons can be made between spectral and imaging observations. And, finally, we have shown that photometric modeling is important for extrapolating specific reflectance calculations to yet unobserved viewing geometries for the purposes of observation planning, either from the ground or from spacecraft platforms.

As discussed earlier, the past decade has seen improvements in the ability of physically motivated photometric models to accurately relate the observed photometric behavior of a regolith-covered body to the physical properties of the surface and the optical properties of its constituents. Studies suggest that, in particular, the Hapke model is realistic when macroscopic surface relief and complex particle shapes can be ignored (Hapke et al., 2009). However the approach is not entirely reliable and work is especially needed to better understand the opposition effect. In particular, the roles of coherent-backscatter, shadow-hiding, and macroscopic surface roughness need to be clearly resolved and tested. Hapke et al. (2012), for example, found no wavelength dependence of the opposition effect for lunar terrains – a result that conflicts with his model for the coherent-backscatter opposition effect (Hapke, 2002; 2012). The relative contributions of CBOE and SHOE have yet to be clearly resolved in photometric studies. When both opposition effects are fit to planetary data, CBOE typically models a very narrow "opposition spike" at phase angles less than a few degrees, while SHOE, which was formerly invoked to model the entire opposition effect now typically describes a broader angular component that may extend to a few tens of degrees in width (cf. Helfenstein et al., 1997). This range overlaps with the predicted phase angle effects from macroscopic surface roughness (Hapke, 1984). It may turn out that the individual contributions of each mechanism cannot be uniquely identified by photometric observation alone, and that polarimetric modeling may provide important missing clues (see Belskaya, et al., this volume). Furthermore, theoretical models have been tested to the point that the viewing geometry conditions necessary for the retrieval of important physical properties (porosity, and average particle phase functions) of the asteroid surface are now defined (Hapke, 2008; Souchon et al., 2011; Helfenstein and Shepard, 2011). These developments should be considered in future investigations, specifically from spacecraft platforms.

Laboratory measurements have been critical to the advancement of the theory and the interpretation of spacecraft measurements (e.g. Helfenstein and Shepard 2011 and references therein).

Finally, spacecraft measurements have been obtained of Ceres, Itokawa, Steins, Lutetia, Annefrank, and Vesta, and all of these data have been fit with photometric models that have deepened our understanding of both asteroid surfaces, and the photometric modeling process. We have reviewed how these additional observations can be compared with the existing observations of Gaspra, Ida, and Eros to form the basis of a developing data set of asteroid light-scattering studies as a function of asteroid spectroscopic classification and mineralogical composition.

Future work should be informed by the need for observations at a wide range of solar phase angles, including near opposition and near or greater than 100º – for the purposes of determining asteroid surface physical characteristics. Future work should also increase the sampling of the spectral classes in observations obtained from spacecraft platforms. Finally, future work should synthesize spacecraft, laboratory, and ground-based surveys of asteroid surface physical properties and their compositional dependence.

Other future work in asteroid photometry might result in sufficient sampling of spatially resolved scattering behavior such that actual physically meaningful scattering model parameters can be mapped across the surface, such as in the work of Sato et al. (2014), who present and interpret Hapke model parameter maps of the lunar surface. These workers find decreased backscattering in the maria relative to the lunar highlands, consistent with known compositional and space weathering differences between maria and highlands, and suggesting powerful new ways to potentially study variations across the surface of an asteroid. For example, the near-future OSIRIS-REx mission will obtain large numbers of repeated observations of the surface of asteroid (101955) Bennu, sampling a wide range of viewing angles for each surface facet, conceivably enabling the first Hapke model parameter maps of an asteroid surface.

**Acknowledgements:** Support for J.-Y. L. is partly from NASA Dawn at Vesta Participating Scientist Program grant NNX13AB82G. P.H. gratefully acknowledges support from NASA Planetary Geology and Geophysics Program grant NNX14AN04G. The thoughtful reviews from Dr. Karri Muinonen and Dr. Bruce Hapke have greatly helped us improve the manuscript.

Table 1 The definitions of some commonly used reflectance quantities

| Quantity | Definition | Formula | Ref* |
|---|---|---|---|
| Bidirectional reflectance | Ratio of the scattered radiance towards $(i, e, \alpha)$ to the collimated incident irradiance | $r(i, e, \alpha) = I(i, e, \alpha)/J$ [ster$^{-1}$] | pp195 |
| Bidirectional reflectance distribution function (BRDF) | Ratio of the scattered radiance towards $(i, e, \alpha)$ to the collimated power incident on a unit area of the surface | $BRDF = I(i, e, \alpha)/J\mu_0 = r/\mu_0$ [ster$^{-1}$] | pp 263 |
| Radiance factor (RADF) | Ratio of the bidirectional reflectance of a surface to that of a perfectly scattering surface$^\$$ illuminated at normal direction | $RADF = \pi r(i, e, \alpha) = [I/F]$ | pp 264 |
| Reflectance factor (or reflectance coefficient, REFF) | Ratio of the reflectance of a surface to that of a perfectly diffused surface under the same conditions of illumination and viewing | $REFF = \pi r/\mu_0 = [I/F]/\mu_0$ | pp 263 |
| Lambertian albedo | Ratio of the total scattered irradiance towards all directions from a Lambert surface to incident power per unit area | $A_L = P_L/J\mu_0$ <br> Perfectly scattering surface has $A_L = 1$ | pp. 187 |
| Normal albedo | Ratio of the reflectance of a surface observed at zero phase angle from an arbitrary direction to that of a perfectly diffuse surface located at the same position, but illuminated and observed perpendicularly | $A_n = \pi r(e, e, 0)$ | pp 296 |
| Geometric albedo (physical albedo) | Ratio of the integral brightness of a body at zero phase angle to the brightness of a perfect Lambert disk of the same size and at the same distance, but illuminated and observed perpendicularly. It is the weighted average of the normal albedo over the illuminated area of the body | $A_p = \int_{2\pi} r(e, e, 0) \mu \, d\Omega$ # | pp. 298 |
| Bond albedo (spherical albedo, or global albedo) | Total fraction of incident irradiance scattered by a body into all directions | $A_s = \dfrac{1}{\pi} \int_{2\pi} \int_{2\pi} r(i, e, \alpha) \mu \, d\Omega_e \, d\Omega_i$ | pp. 301 |
| Bolometric albedo (radiometric albedo) | Average of the spectral albedo $A_s(\lambda)$ weighted by the spectral irradiance of the Sun $J_s(\lambda)$ | $A_b = \dfrac{\int_0^\infty A_s(\lambda) J_s(\lambda) \, d\lambda}{\int_0^\infty J_s(\lambda) \, d\lambda}$ | pp. 302 |
| Phase integral | | $q = 2 \int_0^\pi \Phi_p(\alpha) \sin \alpha \, d\alpha$ | pp. 302 |

Table 2  Commonly used empirical photometric models

| Model | RADF* | Normal Albedo ($A_n$) | Geometric Albedo ($A_p$) | Reference |
|---|---|---|---|---|
| Lambert | $A_L \mu_o f(\alpha)$ | $A_L f(0)$ | $\frac{2}{3} A_L f(0)$ | |
| Lommel-Seeliger | $A_{LS} \frac{\mu_o}{\mu_o + \mu} f(\alpha)$ | $\frac{1}{2} A_{LS} f(0)$ | $\frac{1}{2} A_{LS} f(0)$ | |
| Lunar-Lambert | $A_{LL} \left[ L(\alpha) \frac{2\mu_o}{\mu_o + \mu} + (1 - L(\alpha)) \mu_o \right] f(\alpha)$ $ $L(\alpha) = 1 + A_1 \alpha + A_2 \alpha^2 + A_3 \alpha^3$ | $A_{LL} f(0)$ | $\frac{2 + L(0)}{3} A_{LL} f(0)$ | McEwen (1991; 1996) |
| Minnaert | $A_M \mu_o^{k(\alpha)} \mu^{k(\alpha)-1} f(\alpha)$ $k(\alpha) = k_0 + b\alpha$ # | $A_M f(0)$ | $\frac{2}{2k+1} A_M f(0)$ | Minnaert (1941) Li et al. (2009; 2013b) |
| Akimov | $A_n \cos\frac{\alpha}{2} \cos\left(\frac{\pi}{\pi-\alpha}\left(\Lambda - \frac{\alpha}{2}\right)\right) \frac{\cos^{\frac{\alpha}{\pi-\alpha}} \beta}{\cos \Lambda} f(\alpha)$ | $A_n f(0)$ | $A_n f(0)$ | Shkuratov et al. (2011) |

* $f(\alpha)$ is a surface phase function, generally normalized to unity at zero phase angle.  $A$ with various subscriptions are constants that are directly proportional to the normal albedo and geometric albedo of the surface.  Therefore, the RADF can also be expressed in terms of normal albedo or geometric albedo.
$ $L(\alpha)$ is a partition parameter between the LS term and the Lambert term.  It usually depends on phase angle.
# This is an empirical model of $k(\alpha)$ as adopted by Li et al. (2009; 2013a).  Masoumzadeh et al. (2014) show, from their work on asteroid Lutetia using Rosetta flyby data, that the phase angle dependence of $k$ might be better described with a $2^{nd}$ order polynomial.

Table 3 List of empirical surface phase functions

| Model | Empirical Phase Function | References |
|---|---|---|
| Exponential | $f(\alpha) = e^{\beta\alpha + \gamma\alpha^2 + \delta\alpha^3}$ | Takir et al. (2014) |
| Linear-magnitude | $f(\alpha) = 10^{-0.4\beta\alpha}$ | Li et al. (2009; 2013a) |
| Polynomial-magnitude | $f(\alpha) = 10^{-0.4(\beta\alpha + \gamma\alpha^2 + \delta\alpha^3)}$ | Takir et al. (2014) |
| Lunar/ROLO | $f(\alpha) = C_0 e^{-C_1\alpha} + A_0 + A_1\alpha + A_2\alpha^2 + A_3\alpha^3 + A_4\alpha^4$ | Hillier et al. (1999); Buratti et al. (2011) |
| Linear-Exponential | $f(\alpha) = ae^{-\alpha/d} + b + k\alpha$ | Piironen, 1994; Kaasalainen et al. (2001; 2003) |
| Akimov | $f(\alpha) = \frac{e^{-\mu_1\alpha} + me^{-\mu_2\alpha}}{1+m}$ [*] | Akimov (1988) |

[*] $\mu_1$ and $\mu_2$ are model parameters, not to be confused with $\mu_0$ and $\mu$, which are the cosines of incidence angle and emission angle, respectively.

Table 4 Disk-integrated phase functions for various commonly used photometric models

| Model | Disk-Integrated Phase Function[*] | References |
|---|---|---|
| Lambert | $\Phi(\alpha) = [\sin\alpha + (\pi - \alpha)\cos\alpha]\frac{f(\alpha)}{\pi f(0)}$ | Hapke (2012b) pp. 107, Eq. 6.13 |
| Lommel-Seeliger | $\Phi(\alpha) = \left(1 - \sin\frac{\alpha}{2}\tan\frac{\alpha}{2}\ln\cot\frac{\alpha}{4}\right)\frac{f(\alpha)}{f(0)}$ | Hapke (2012b) pp. 108, Eq. 6.14 |
| Lunar-Lambert | $\Phi(\alpha) = \frac{\frac{2}{3\pi}[1-L(0)][\sin\alpha + (\pi-\alpha)\cos\alpha] + L(0)f(\alpha)\left(1-\sin\frac{\alpha}{2}\tan\frac{\alpha}{2}\ln\cot\frac{\alpha}{4}\right)}{\frac{2}{3}[1-L(0)] + L(0)f(0)}$ | Buratti and Veverka (1983) |
| Hapke | $\Phi(\alpha) = \frac{r_0}{2A_p}\left\{\left[\frac{(1+\gamma)^2}{4}\{[1+B_{S0}B_S(\alpha)]p(\alpha) - 1\} + [1-r_0]\right]\left(1 - \sin\frac{\alpha}{2}\tan\frac{\alpha}{2}\ln\cot\frac{\alpha}{4}\right) + \frac{4}{3}r_0\frac{\sin\alpha + (\pi-\alpha)\cos\alpha}{\pi}\right\}[1 + B_{C0}B_C(\alpha)]$ | pp. 300, Eq. 11.42 |

[*] All phase functions are normalized to unity at zero degree phase angle.

Table 5 Particle single scattering phase functions

| Name | Expression | Parameters | Reference |
|---|---|---|---|
| Modified Schoenberg | $P(\alpha) = [\sin|\alpha|+(\pi-|\alpha|) \cos(|\alpha|)/\pi + 0.1(1-\cos(|\alpha|))^2]$ | | Hapke (1963,1966) |
| Two-Parameter Legendre Polynomial (2PLP) | $P(\alpha) = 1 + b \cos(\alpha) + c \left(\frac{3}{2}\cos^2(\alpha) - \frac{1}{2}\right)$ | $b,c$ | Hapke (1981) |
| One-Parameter Henyey-Greenstein (1PHG) | $P(\alpha) = (1-g^2)/(1+ 2g \cos(\alpha) + g^2)^{3/2}$ | $g$ | Buratti and Veverka (1983) |
| Two-Parameter Henyey-Greenstein (2PHG, form #1) | $P(\alpha) = \frac{(1+c)}{2}\frac{(1-b^2)}{(1-2b\cos(\alpha)+b^2)^{3/2}} + \frac{(1-c)}{2}\frac{(1-b^2)}{(1+2b\cos(\alpha)+b^2)^{3/2}}$ | $b,c$ | McGuire and Hapke (1995) |
| Two-Parameter Henyey-Greenstein (2PHG, form #2) | $P(\alpha) = (1-c)\frac{(1-b^2)}{(1+2b\cos(\alpha)+b^2)^{3/2}} + c\frac{(1-b^2)}{(1-2b\cos(\alpha)+b^2)^{3/2}}$ | $b,c$ | Hartman and Domingue (1998) |
| Three-Parameter Henyey-Greenstein (3PHG, form #1) | $P(\alpha) = (1-f)\frac{(1-g_1^2)}{(1+2g_1\cos(\alpha)+g_1^2)^{3/2}} + (f)\frac{(1-g_2^2)}{(1+2g_2\cos(\alpha)+g_2^2)^{3/2}}$ | $g_1,g_2,f$ | Helfenstein et al. (1991) |
| Three-Parameter Henyey-Greenstein (3PHG, form #2) | $P(\alpha) = (f)\frac{(1-g_1^2)}{(1+2g_1\cos(\alpha)+g_1^2)^{3/2}} + (1-f)\frac{(1-g_2^2)}{(1-2g_2\cos(\alpha)+g_2^2)^{3/2}}$ | $g_1,g_2,f$ | Deau and Helfenstein (2015) |
| Lumme-Bowell | $P(\alpha) = 0.95e^{-0.4\alpha} + 16.15\, e^{-4.0\alpha}$ | | Lumme and Bowell (1981b) |

Table 6  Comparison of the Hapke photometric parameters of asteroids [*]

| Object | Type | $\varpi$ (visible) | $h$ | $B_0$ | $g$ | $\theta$ (°) | Reference |
|---|---|---|---|---|---|---|---|
| Average S | | 0.23 | 0.08 | 1.6 | -0.27 | 20 | Helfenstein and Veverka (1989) |
| Average C | | 0.037 | 0.025 | 1.03 | -0.47 | 20 | Helfenstein and Veverka (1989) |
| Average V (NEOs and Vesta) | | 0.51 | 0.098 | 1.0 | -0.26 | 32 | Hicks et al. (2014) |
| (4) Vesta | V | 0.51 | 0.07 | 1.7 | -0.24 | 18 | Li et al. (2013a) |
| (951) Gaspra | S | 0.36 | 0.06 | 1.63 | -0.18 | 29 | Helfenstein et al (1994) |
| (243) Ida | S | 0.22 | 0.02 | 1.53 | -0.33 | 18 | Helfenstein et al (1996) |
| Dactyl | S | 0.21 | (0.020) | (1.53) | -0.33 | 23 | Helfenstein et al. (1996) |
| (433) Eros | S | 0.43 | 0.022 | 1.0 | -0.29 | 28 | Li et al. 2004 |
| (25143) Itokawa | S | 0.36 | (0.022) | (1.0) | -0.51 | (20) | Lederer et al. (2005) |
| (25143) Itokawa | S | 0.42 | 0.01 | 0.87 | -0.35 | 26 | Kitazato et al. (2008) |
| (5535) Annefrank | S | 0.41 | 0.015 | 1.32 | -0.19 | 20 | Hillier et al. (2011) |
| (1862) Apollo | S | | | | | | Helfenstein and Veverka (1989) |
| (253) Mathilde | C | 0.034 | 0.094 | 3.18 | -0.27 (2-term fit) | 25 | Clark et al. (1999) |
| (1) Ceres | C | 0.070 | 0.06 | 1.6 | -0.4 | 44 | Helfenstein and Veverka (1989); Li et al. (2006) |
| (2867) Šteins | E | 0.57 | 0.062 | 0.60 | -0.30 | 28 | Spjuth et al. (2012) |
| (21) Lutetia | M | 0.23 | 0.044 | 1.93 | -0.25 | 25 | Masoumzadeh et al. (2014) |

[*] Values inside parentheses are assumed.

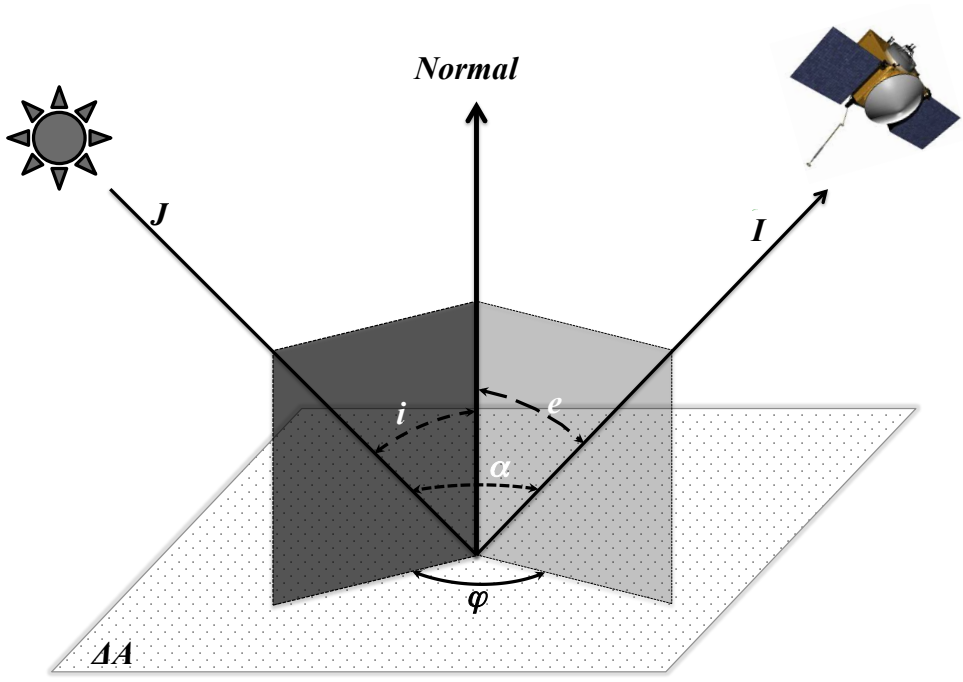

Fig. 1. Schematic diagram of bidirectional reflectance from a surface element $\Delta A$, showing the various angles. The plane containing $J$ and $I$ is the scattering plane. If the scattering plane also contains $N$, it is called the principal plane. $\psi$ is the azimuthal angle between the planes of incidence and emission [$\cos(\alpha) = \cos(i)\cos(e)+\sin(i)\sin(e)\cos(\psi)$].

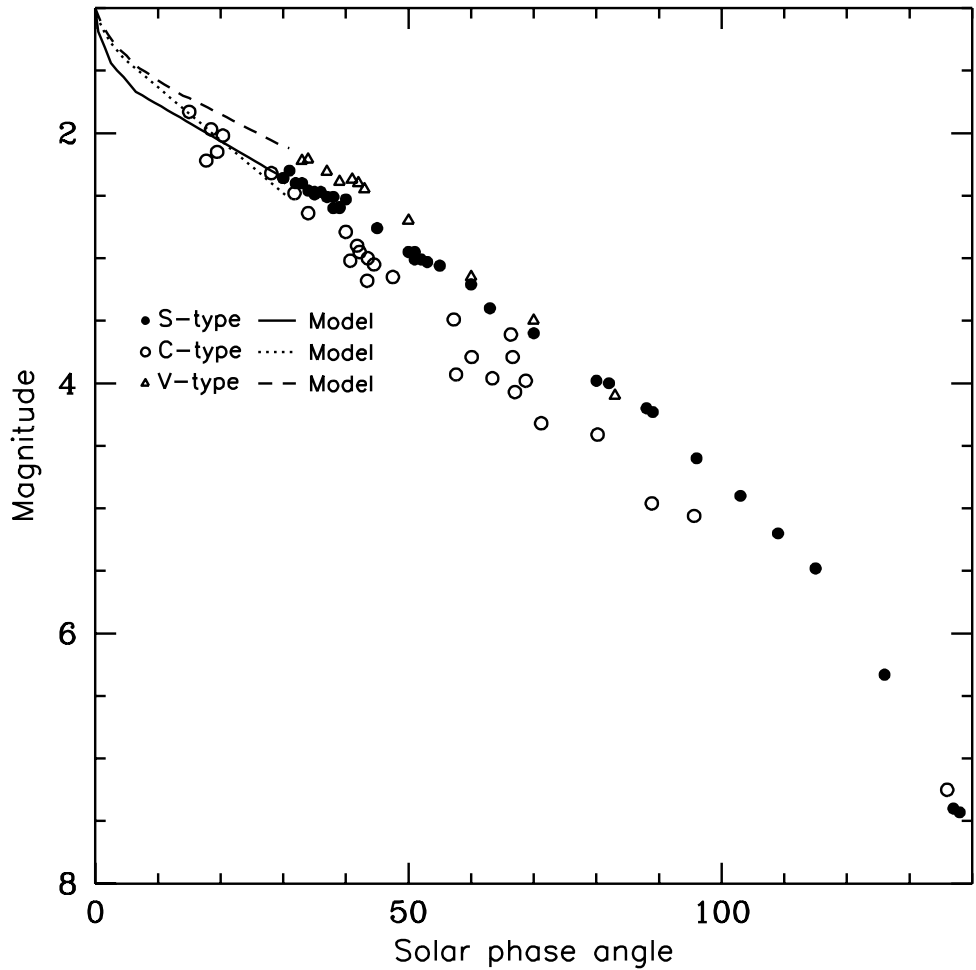

Fig. 2. The phase functions of three asteroid classes, all normalized at zero degree phase angle. The lines at phase angles less than ~30° are the respective best-fit models using the data from Helfenstein and Veverka (1989) for S and C types (*Asteroids II* book), and a composite Vesta curve from Hicks et al. (2014) for V types. The various symbols are from actual measurements of asteroids listed in Table 8. Data from Mathilde (Clark et al., 1999) and Bennu (B-type) (Takir et al., 2014) are used for additional C types beyond 30° phase angle. Data for Gaspra (Helfenstein et al., 1994), Lutetia (Masoumzadeh et al., 2014), and Eros (Li et al., 2004) are used for S type beyond 30° phase angle.

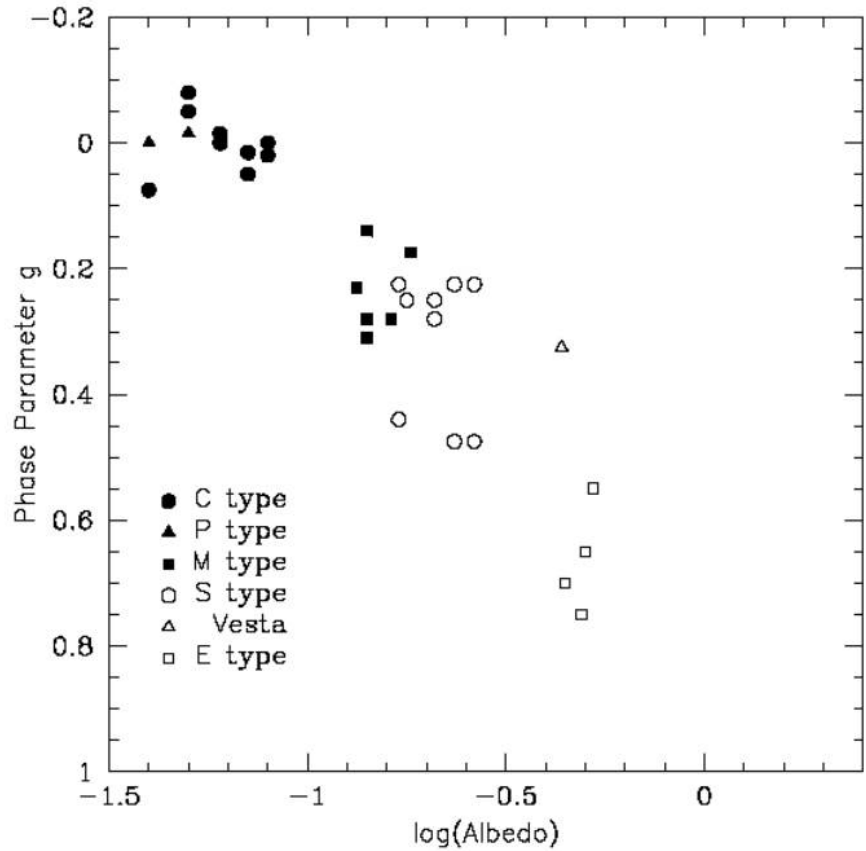

Fig. 3. The H and G parameters of the Lumme-Bowell model correlate well with asteroid class. The low-albedo primitive type have lower phase parameters G, while the higher albedo asteroids have larger G. (based on Belskaya and Shevchenko, 2000; Hicks et al. 2014)